\documentclass{article}
\usepackage{fix-cm}

\usepackage{PRIMEarxiv}

\usepackage[utf8]{inputenc} 
\usepackage[T1]{fontenc}    
\usepackage{lmodern}        
\usepackage{textcomp}      
\usepackage{hyperref}       
\usepackage{url}            
\usepackage{booktabs}       
\usepackage{amsfonts}       
\usepackage{nicefrac}       
\usepackage{microtype}      
\usepackage{lipsum}
\usepackage{fancyhdr}       
\usepackage{graphicx}   
\usepackage{amsmath}
\usepackage{textgreek}
\usepackage{hyphenat} 
\graphicspath{{media/}}     

\title{Associating Trajectories with Quantum Processes by Equivalent Spectra}

\author{
  Sriram Aditya Gundla \\
  Affiliation: 10x Accel\\West windsor-Plainsboro High School North\\Plainsboro Township, NJ
   \And
   Arunabh Pratik \\
  Affiliation: 10x Accel \\
  Princeton International School of Mathematics and Science (PRISMS)\\
  Princeton, NJ\\
  \And 
  Shyla Rathore\\Affiliation: 10x Accel\\John P. Stevens High School\\Edison, NJ
  \And Dr. Elliott Tammaro\\10x Accel mentor\\Chestnut Hill College\\ Philadelphia, PA}

\begin{document}
\maketitle

\begin{abstract}
Quantum mechanics abandoned the classical notion of a particle trajectory, yet trajectories remain conceptually appealing for resolving foundational issues in quantum mechanics. Bohmian mechanics offers one route to associating trajectories with quantum processes, but it requires explicit non-locality to remain consistent with EPR-type experiments, placing it in tension with special relativity. We propose an alternative approach: rather than deriving trajectories from a quantum guidance equation, we search for classical trajectories whose radiated frequency spectra match those of quantum processes such as atomic transitions. Using the Liénard-Wiechert potentials, we compute the electric and magnetic fields generated by a point charge along a given trajectory, numerically solving the retarded-time and from these fields obtain the power spectrum of emitted radiation. By fitting the parameters of a chosen family of trajectories, we identify "equivalent spectrum" trajectories that reproduce a target frequency distribution, including ones of quantum mechanical origin. We discuss the implications of this method and propose future work to determine whether equivalent spectrum trajectories belong to a family governed by a differential equation, which would constitute a quantum-equivalent equation of motion.

\end{abstract}

\keywords{Quantum mechanics\and Hidden variable theory\and Atomic Spectra\and Li\'enard-Wiechert Potentials}

\section{Introduction}
Quantum mechanics was born, in part, out of necessity for the description of atomic emission spectra. The fundamental difficulties in modeling the atom had been established. Sir Joseph Larmor, Alfred-Marie Li\'enard, and Emil Wiechert \cite{larmor1897,lienard1898,wiechert1901,jackson1999}, using Maxwell's equations, calculated that accelerated charges emit radiation so that the frequency spectrum of the emitted light can be written in terms of the Fourier decomposition of the trajectory. Ernst Rutherford and others demonstrated that the electrons are in bound states outside of the positively charged nucleus\cite{rutherford1911,bohr1913}. It would thus be expected through classical reasoning that the electron, being accelerated by the Coulombic force from the nucleus, would radiate. This, in turn, would cause orbital decay and eventually the electron would spiral into the nucleus\cite{bohr1913}. Classically, the atom would be unstable. This is a major deficiency of the classical approach and necessitated the development of quantum mechanics.   

Outside of the question of long term orbital stability (the electron does not radiate in the ground state) the description provided by the classical approach is qualitatively, and in certain regards, quantitatively, similar to what is observed in the phenomena of atomic spectra. Electrons in the so-called excited states do radiate, and they exhibit a kind of orbital decay until reaching the ground state. Emission spectra are composed of discrete peaks, which is consistent with the frequency distribution (spectrum) of a charge following a multiperiodic trajectory.  

Trajectories remain essential in the old quantum theory, which uses Bohr-Sommerfeld quantization \cite{bohr1913,sommerfeld1916}, but with the development of matrix mechanics, the earliest formulation of quantum mechanics, the trajectory had to be abandoned \cite{heisenberg1925,born1926}. While quantum mechanics is exceedingly successful, it would be compelling to preserve the notion of a trajectory. Maintaining the existence of a trajectory may partially resolve, or even eliminate, the measurement problem \cite{bell1987}. Moreover, through technological developments quantum mechanics is becoming applicable to physically larger systems. For example, researchers at Swiss Federal Institute of Technology placed a sapphire crystal with a mass of 16 micrograms in a quantum superposition of two oppositely directed vibrational states\cite{chan2021}. This is incredible because the crystal is large enough to be seen with the naked eye. Of course it was not a superposition of macroscopically distinct position states, but if quantum mechanics does not break sooner then there will come an experiment such that a large system is detected at A and again at B while otherwise being in a superposition of macroscopically distinct position states. In such experiments there is increasing tension between classical and quantum mechanics. Experiments could also be explained in a more intuitive fashion, and the traditional explanation of many experiments that utilize the (often assumed classical) movement of atoms of molecules would be better justified.  

In this paper we report on our group’s attempt to associate a trajectory with quantum processes. The association of a trajectory with a quantum process is not new. It is firmly established that Bohmian mechanics accomplishes precisely this\cite{bohm1952a,bohm1952b,holland1993}. However, there are some well known issues with Bohmian mechanics. Quantum mechanics correctly describes the kind of non-locality that is observed in nature through entanglement and other EPR-type experiments\cite{einstein1935,aspect1982}. In Bohmian mechanics this non-locality must be made explicit and as a result it is mathematically at odds with the special theory of relativity. As a result, Bohmian field theory, and therefore the Bohmian trajectories of atomic electrons in quantum jumps, is an area of active research. Instead, we will proceed through the following methodology. Given the trajectory of a point charge, such as an electron, the electric and magnetic fields may be calculated directly from the trajectory via the Lienard-Wiechert potentials, \begin{equation}\phi(\mathbf{r},t)
=
\frac{1}{4\pi\varepsilon_0}
\left.
\frac{q}
{(1-\mathbf{n}\cdot\boldsymbol{\beta})\,R}
\right|_{\mathrm{ret}}\end{equation}
\begin{equation}
\mathbf{A}(\mathbf{r},t)
=
\frac{\mu_0}{4\pi}
\left.
\frac{q\,\mathbf{v}}
{(1-\mathbf{n}\cdot\boldsymbol{\beta})\,R}
\right|_{\mathrm{ret}},\end{equation} from which the electric and magnetic fields may be calculated. \begin{equation}\mathbf{E}
=
-\nabla\phi
-\frac{\partial\mathbf{A}}{\partial t}\end{equation} 
\begin{equation}\mathbf{B}
=
\nabla\times\mathbf{A}.\end{equation}
   From the electric and magnetic fields the frequency distribution (spectrum) follows \cite{lienard1898,wiechert1901,jackson1999}. 
   
   Many quantum processes, atomic transition spectra for example, are characterized by an associated frequency distribution. One may ask, given a frequency distribution which trajectory, if any, can reproduce this? It may be said that the association between quantum processes and trajectories that is considered here is through a kind of “reverse engineering.” The trajectories are intended to be deduced from the frequency distribution rather than the frequency distribution being calculated from a known trajectory.

\section{Methodology}
\fontsize{11}{13}\selectfont

Given the trajectory of charged point particle the Li\'enard-Wiechert potentials allow one to calculate the electric and magnetic fields, and from these the frequency distribution of any emitted radiation may be computed. A difficulty that arises is that the potentials are evaluated at the retarded time $t_{ret}$, which satisfies \begin{equation}t_{\rm ret}
=
t
-
\frac{
\left|\mathbf{r}-\mathbf{r}_0(t_{\rm ret})\right|
}{c}, \label{retarded time}\end{equation}
where $\mathbf{r}_0(t)$ is the particle trajectory \cite{larmor1897,lienard1898,wiechert1901,jackson1999}.
Only for the simplest trajectories can equation \ref{retarded time} be solved manually. We wrote programs in both C++ and Python that solved the retarded time equation for a grid of points and times. That grid of solutions was then used to compute the electric and magnetic fields on grid. We are mainly interested in the radiation fields, which satisfy $\mathbf{B} = \frac{1}{c}\mathbf{n}\times \mathbf{E}$ so the power per unit frequency distribution is 
\begin{equation}P(\omega)\propto |\mathbf{\tilde{E}}(\omega)|^2,\end{equation} where \begin{equation}\mathbf{\tilde{E}}(\omega)=\int_{-\infty}^{+\infty}\mathbf{E}(t)e^{i \omega t}dt.\end{equation} The grid of solutions to equation \ref{retarded time} was used to compute $P(\omega)$ on grid \cite{jackson1999}. This code is available from the authors upon request.

Suppose a frequency distribution $\mathcal{F}(\omega)$ is provided along with a family of trajectories, depending on parameters $(\alpha_1, \alpha_2,...\alpha_n)$. From this parameterized family of trajectories the frequency distribution can be computed, which will now depend on the parameters $(\alpha _1,...\alpha_n)$. A best fit of the frequency can be made by adjusting the parameters to match the given frequency distribution $\mathcal{F}(\omega)$. This method may be employed even if $\mathcal{F}(\omega)$ has quantum mechanical origins. Thus a trajectory may be associated with quantum mechanical process because of the equivalence of their spectra. 
\section{Discussion and Future Work}
Associating a quantum process with a trajectory is well-established through Bohmian mechanics, however Bohmian mechanics is problematic. In order to be consistent with the non-locality observed in nature through EPR-type experiments it must be non-local, but Bohmian mechanics is at a disadvantage because it has explicit non-locality. Bohmian mechanics can, in principle, handle electron transitions and even other phenomena that require QED, it is essentially reduced to solving the quantum field theoretic problem and then solving for a trajectory from the guidance equation. The approach that we followed herein is to associate trajectories that generate the same spectrum as quantum processes, which we refer to as ``equivalent spectrum'' trajectories. This method opens an interesting path for future work. Are the equivalent spectrum trajectories part of family that solves a differential equation? If this can be answered in the affirmative then this differential equation would be a quantum-equivalent equation of motion.

\bibliographystyle{unsrt}  
\bibliography{references}

\end{document}